\documentclass[sigplan]{acmart}

\usepackage{graphicx}
\usepackage{paper_alcides}
\usepackage{macros_g}
\usepackage{url}
\usepackage{cleveref}
\usepackage{float}
\usepackage{caption}
\usepackage{listings}

\newfloat{lstfloat}{t}{lop}
\floatname{lstfloat}{Listing}

\usepackage[utf8]{inputenc}

\setcopyright{acmcopyright}
\copyrightyear{2018}
\acmYear{2018}
\acmDOI{XXXXXXX.XXXXXXX}

\acmConference[Conference acronym 'XX]{Make sure to enter the correct
  conference title from your rights confirmation emai}{June 03--05,
  2018}{Woodstock, NY}
\acmPrice{15.00}
\acmISBN{978-1-4503-XXXX-X/18/06}

\begin{document}

\title[Data types as a more ergonomic frontend for GGGP]{Data types as a more ergonomic frontend for Grammar-Guided Genetic Programming}


\author{Guilherme Espada}
\email{gjespada@ciencias.ulisboa.pt}
\orcid{0000-0001-8128-7397}
\affiliation{%
  \institution{LASIGE, Faculdade de Ciências da Universidade de Lisboa}
  \country{Portugal}
}

\author{Leon Ingelse}
\email{lingelse@lasige.di.fc.ul.pt}
\orcid{0000-0001-6067-6318}
\affiliation{%
  \institution{LASIGE, Faculdade de Ciências da Universidade de Lisboa}
  \country{Portugal}
}

\author{Paulo Canelas}
\email{pacsantos@ciencias.ulisboa.pt}
\orcid{0000-0002-0154-8989}
\affiliation{%
  \institution{School of Computer Science \\ Carnegie Mellon University}
  \country{USA}
}
\affiliation{%
  \institution{LASIGE, Faculdade de Ciências da Universidade de Lisboa}
  \country{Portugal}
}

\author{Pedro Barbosa}
\email{psbarbosa@ciencias.ulisboa.pt}
\orcid{0000-0002-3892-7640}
\affiliation{%
  \institution{LASIGE, Faculdade de Ciências da Universidade de Lisboa}
  \country{Portugal}
}
\affiliation{%
  \institution{Instituto de Medicina Molecular}
  \country{Portugal}
}

\author{Alcides Fonseca}
\email{alcides@ciencias.ulisboa.pt}
\orcid{0000-0002-0879-4015}
\affiliation{%
  \institution{LASIGE, Faculdade de Ciências da Universidade de Lisboa}
  \country{Portugal}
}

\renewcommand{\shortauthors}{Espada et al.}

\begin{abstract}

Genetic Programming (GP) is an heuristic method that can be applied to many Machine Learning, Optimization and Engineering problems. In particular, it has been widely used in Software Engineering for Test-case generation, Program Synthesis and Improvement of Software (GI).

Grammar-Guided Genetic Programming (GGGP) approaches allow the user to refine the domain of valid program solutions. Backus Normal Form is the most popular interface for describing Context-Free Grammars (CFG) for GGGP. BNF and its derivatives have the disadvantage of interleaving the grammar language and the target language of the program.

We propose to embed the grammar as an internal Domain-Specific Language in the host language of the framework. This approach has the same expressive power as BNF and EBNF while using the host language type-system to take advantage of all the existing tooling: linters, formatters, type-checkers, autocomplete, and legacy code support. These tools have a practical utility in designing software in general, and GP systems in particular.

We also present Meta-Handlers, user-defined overrides of the tree-generation system. This technique extends our object-oriented encoding with more practicability and expressive power than existing CFG approaches, achieving the same expressive power of Attribute Grammars, but without the grammar vs target language duality.

Furthermore, we evidence that this approach is feasible, showing an example Python implementation as proof. We also compare our approach against textual BNF-representations w.r.t. expressive power and ergonomics. These advantages do not come at the cost of performance, as shown by our empirical evaluation on 5 benchmarks of our example implementation against PonyGE2. We conclude that our approach has better ergonomics with the same expressive power and performance of textual BNF-based grammar encodings.
\end{abstract}

\ccsdesc[500]{Computer systems organization~Embedded systems}
\ccsdesc[300]{Computer systems organization~Redundancy}
\ccsdesc{Computer systems organization~Robotics}
\ccsdesc[100]{Networks~Network reliability}

\keywords{Grammar-guided Genetic Programming, Strongly-Typed Genetic Programming, Genetic Programming Framework}
\maketitle

\section{Introduction}


Genetic Programming (GP) has been applied to several domains~\cite{DBLP:journals/itiis/AhvanooeyLWW19}, including test-case generation~\cite{DBLP:conf/gecco/EmerV02}, program synthesis~\cite{DBLP:conf/eurogp/ForstenlechnerF17} and automatic optimization of programs~\cite{DBLP:conf/synasc/Langdon14}. Being a search-based method, GP often requires the user to encode domain-specific restrictions to the possible solutions.

Throughout the Genetic Programming procedure, many solutions (also called programs or trees) are generated, modified and combined in order to search the one that has the best performance or solves the problem. One of the main challenges of using GP is to define the boundaries of the search-space in a way that is performant and efficient. Encoding domain-knowledge in the definition of the search space is desirable, so many candidate, but invalid solutions are removed from the search.Grammar-Guided and Strongly-Typed Genetic Programming (GGGP~\cite{whigham1995Grammars} and STGP~\cite{DBLP:journals/ec/Montana95}) are two popular families of approaches to guide the random solution generation by the user. To restrict valid solutions, GGGP relies on user-defined grammars while STGP relies on user-defined type annotations.

Within each family of approaches, there are different trade-offs between expressive power and computational complexity. Context-Free Grammars~\cite{whigham1995Grammars} can be used with little overhead in processing, while Attribute Grammars~\cite{DBLP:conf/iwinac/CruzPA05} can encode more complex restrictions. Similarly, polymorphic type systems~\cite{DBLP:conf/eurogp/Yu01} are more expressive than monomorphic ones, but they have a higher overhead during the evolutionary process.

However, in practice, popular implementations of these systems are limited to Context-Free grammars (CFG) (e.g., PonyGE2~\cite{DBLP:conf/gecco/FentonMFFHO17}) and base types (e.g., DEAP~\cite{DBLP:journals/jmlr/FortinRGPG12} and Jenetics~\cite{wilhelmstotter2017jenetics}).

\subsection*{Problem Statement}

GGGP requires users to express the search-space of valid programs in the Backus Normal Form (BNF) or one of its derivations, such as the Extended Backus Normal Form (EBNF). We consider BNF and its derivatives to be a textual editing medium of the target language, as concepts in the target language (like types, functions, literals) have no meaning in the grammar language. As such, users are treating the target language as character streams, instead of treating them as semantic code units, as they do in code editors.

The BNF approach is not ergonomic nor error-prone due to the interleaving of two languages: the BNF meta-syntax and the syntax of the target language, i.e., the language of the generated programs. As an example, \Cref{code:mismatch} depicts two common errors that both novices and experts do while writing grammars. These are not detected immediately, and are typically only revealed during execution.

\begin{lstfloat}
\begin{lstlisting}[language=Python,caption={Example of a grammar with two possible bugs: the \texttt{np} module may not be imported, and the $+$ operator may be applied to a string and integer, causing a crash.},label=code:mismatch]
<S> ::= <S> "+" <S> | "x" | 1 | np.array([<A>])
<A> ::= [0-9]+,<A> | [0-9]+
\end{lstlisting}
\end{lstfloat}

Ideally, these two bugs would be detected immediately in most development environments: the type-checker would identify a missing import, and the type mismatch in the operator. Is it really necessary that, to implement a grammar, developers have to lose access to an array of tooling that has shown to be useful in software engineering? Or is there a more ergonomic interface that is integrated into the host language ecosystem, providing access to code editors, syntax highlighting, linting, formatting, test suites, etc...?

If there is, such an approach could reduce the barrier to adoption of GGGP, STGP, and similar systems, increasing the productivity of practitioners and reducing the number of possible bugs that are not caught earlier due the mismatch between the host language and the grammar language.

It is important to notice that usability has been identified as a relevant open challenge for GP both in 2010~\cite{DBLP:journals/gpem/ONeillVGB10}, and again in 2020, when it was identified as one of the 10 major challenges for the following 15 years~\cite{DBLP:journals/gpem/SipperM20}. This work addresses exactly that problem, especially in the case where practitioners encode domain information to improve the performance of GP.

\subsection*{Approach}

Our first contribution is an object-oriented encoding of grammars to be used in GGGP, regardless of representation. In short, non-terminals are represented as abstract classes (or interfaces, or traits) and productions are represented as concrete classes that implement the corresponding abstract class.

As a running example, we will use grammars to describe the VectorialGP example in \citet{DBLP:conf/eurogp/AzzaliVSBG19}. \Cref{code:gggp_ex} describes that subset using a PonyGE2's flavor of BNF, while \Cref{code:stgp_ex} encodes the same grammar using the proposed approach, both implemented in Python.

\begin{lstfloat}
\begin{lstlisting}[language=Python,caption={PonyGE2 grammar for a subset of VectorialGP},label=code:gggp_ex]
<Real> ::= <RealLit> | DATASET[GE_RANGE:3] | <RealLit> + <RealLit> | np.average(<Vector>)
<RealLit> ::= <d>.<d>
<d> ::= GE_RANGE:10<d> | 
<Vector> ::= DATASET[3 + GE_RANGE:3]
\end{lstlisting}
\end{lstfloat}

\begin{lstfloat}
\begin{lstlisting}[language=Python,caption={Proposed Object-Oriented encoding of grammar in \cref{code:gggp_ex}.},label=code:stgp_ex]
@abstract class Real: pass
@abstract class Vector: pass

class FloatLiteral(Real):
	value: float	
	def evaluate(self):
		return self.value
		
class RealFeature(Real): # First three columns have reals
	column: Annotated[int, IntBetween[0,3]] 
	def evaluate(self):
		return DATASET[self.column]

class RealSum(Real):
	left: Real, right: Real
	def evaluate(self):
		return self.left.evaluate() + self.right.evaluate()
	
class Average(Real):
	values: Vector
	def evaluate(self):
		return np.average(self.values)

class VectorFeature(Vector): # Last three columns have vectors
	column: Annotated[int, IntBetween[3,6]] 
	def evaluate(self):
		return DATASET[self.column]
\end{lstlisting}
\end{lstfloat}

There is a direct correspondence between the abstract and concrete types, and the non-terminals and productions of the grammar. This direct correspondence is an advantage of this approach, that does not add more cognitive overhead when using the Object-Oriented encoding.

Our proposal is also more verbose (visible in the number of lines of code, which tells nothing about code complexity~\cite{DBLP:conf/metrics/Rosenberg97}), but is more ergonomic, when measured using Kiper's framework~\cite{DBLP:journals/jsi/Kiper94}. In particular, developers can have immediate feedback about their grammars and semantic actions using the target language tooling like type-checkers, autocompleters, linting and test frameworks.
It is important to notice that our user-facing encoding of grammars supports any internal representation, from Whigham's CFG-GP~\cite{whigham1995Grammars} to Grammatical Evolution~\cite{Ryan1998GE}, regardless of which one is better in a given problem~\cite{whigham2015examining,lourenco2017comparitive,ingelse2022benchmarking}.

Our second contribution, that builds on top of the first, is the usage of Meta-Handlers. Meta-Handlers are type annotations that can be used to efficiently override the tree-generation process, to achieve a higher (formal and practical) expressive power when compared to BNF and its extensions.

We will now present the related work (\cref{sec:relatedword}), describe our approach (\cref{sec:approach})
, present our evaluation (\cref{sec:evaluation}) and draw conclusions (\cref{sec:conclusions}).

\section{Related Work}
\label{sec:relatedword}

\subsection{Search-space Restriction}

The benefits of GP are sometimes hampered by very large or unbounded search spaces \cite{DBLP:books/daglib/0006025}. The search space can be restricted using grammars (GGGP~\cite{whigham1996searchGGGP}) or using types (STGP~\cite{DBLP:journals/ec/Montana95}), to introduce bias in the population.

The initial proposal of GGGP used context-free grammars (CFG) in the Backus Normal Form to generate trees, and is referred to as CFG-GP~\cite{whigham1995Grammars}. \citet{Ryan1998GE} proposed Grammatical Evolution, using variable-length string genomes to represent programs, with the grammar being used in the genotype-phenotype mapping. \citet{DBLP:conf/eurogp/RyanNO02} introduced Attribute Grammars as a more expressive alternative to CFG grammars, albeit more computationally expensive. 

Strongly-Typed Genetic Programming~\cite{DBLP:journals/ec/Montana95} only generates trees that are valid by construction, showing success in practice~\cite{DBLP:conf/icga/HaynesWSS95,DBLP:conf/epia/Pinheiro0V15}.
Refinement-Typed GP~\cite{DBLP:conf/ppsn/FonsecaSS20} (RTGP) extends STGP with dependent types, increasing the expressive power of the system.
Strongly-Formed GP~\cite{DBLP:conf/eurogp/CastleJ12} (SFGP) also extends STGP with the ability to distinguish node-to-node from node-to-terminal relations, allowing for higher specificity and thus a reduction of the search space.

\subsection{Genetic Programming frameworks}

While GE libraries exist for C (libGE~\cite{Nicolau2006libGE}), Java (GEVA~\cite{oneill2008geva}) and R (gramEvol~\cite{noorian2016gramevol}), PonyGE2~\cite{DBLP:conf/gecco/FentonMFFHO17} in Python is the most mature framework. In PonyGE2, the user is required to write a grammar in BNF notation and to implement the fitness function in Python. We have identified a few limitations of PonyGE2: It is a framework, and not a library that you can use within other programs; it requires the user to be familiar with BNF notation; and it does not support code completion, checking, or linting inside the grammar.

Different frameworks support restricting the shape of the tree with STGP: DEAP~\cite{DBLP:journals/jmlr/FortinRGPG12}, PyStep~\cite{khourypystep} and \texttt{monkeys}~\cite{monkeys} (Python), Jenetics~\cite{wilhelmstotter2017jenetics} (Java) and RGP~\cite{DBLP:conf/gecco/FlaschMB10} (R).
Except for \texttt{monkeys}, other libraries are limited to basic types and functions. While this option may be more efficient, it limits the expressive power of the solution, as an AST cannot be easily created and manipulated. On the other hand, \texttt{monkeys} supports using the internal AST of Python to generate Python code. This framework requires the user to know the internal representation of Python, a very complex and large language. Furthermore, \texttt{monkeys} does not support custom generators for code, and does not use the default type annotations in Python, requiring custom annotations that can upset usage of pre-existing code.

\section{The Data Type Approach}
\label{sec:approach}

Our approach relies on two main concepts: Type Hierarchies as Grammars, and Meta-Handlers as user-customizable overrides for tree generation. The first concept provides the same expressive power as BNF in a more ergonomic interface. The second allows users of GGGP frameworks to encode even more expressive constraints on the grammar.

\subsection{Type-Hierarchy Encoding of Grammars}
\label{subsec:types_as_grammars}

Grammar-Guided approaches require the practitioner to restrict the grammar of valid, candidate programs. The grammar is typically described using BNF or one of its variants. Each grammar describes productions from non-terminal symbols to a sequence of terminal and non-terminal symbols. Because terminal symbols belong to the target programming language, productions mix the target language with the meta-language (BNF). As such, there is no tooling to verify that a grammar generates valid programs in the target language. This can be checked by generating individuals and evaluating them, but it is not complete. It may take many generations before generating an individual that is not valid, and the user finds that the grammar was not properly designed.

STGP is less error-prone because syntactic and semantic errors in the specification of the search space is detected by type-checkers. If the user selects the wrong type for a function, the program will not compile.

Our approach combines the expressive power of GGGP with the ergonomic advantages of STGP. Briefly, a CFG grammar is encoded using the type system of the target language, with the grammar being specified in the same programming language that is being generated.  Thus, all existing tooling can be used and applied to the grammar definition, and it avoids mixing different syntaxes in the same file.

Formally, a grammar $G$ is defined as the quadruple $<N,\Sigma,P,S>$, where $N$ is the set of non-terminals, $\Sigma$ is the set of terminal symbols, $P$ is the set of productions in the grammar and $S$ is the starting symbol to produce the grammar. Most often, non-terminal ($N$) and terminal ($\Sigma$) symbols are distinguished using a special syntax (e.g., enclosing the non-terminal symbol with angled brackeds, like in \texttt{<Real>}), and the starting symbol is often identified by being the first one defined, or having a pre-defined name, such as \texttt{start} or \texttt{S}. Hence, the main concern when defining a grammar is defining the production and the non-terminal symbols.

\textbf{Rule 1:} Non-terminal symbols are defined as abstract data types, interfaces, abstract classes, traits, protocols or type-classes. The distinction between them~\cite{DBLP:conf/rex/Cook90} is not relevant and can be chosen according to the preference of the implementer. Querying the existing abstract types builds the non-terminal set.

\textbf{Rule 2:} Each grammar production will correspond to a different concrete data type that implements (or inherits from) the corresponding non-terminal abstract type. The main requirement is that this relationship can be queried by the GP implementation. Querying the relationship between concrete types and their abstract types is what builds the production set of the grammar.

\textbf{Rule 3:} Terminal symbols exist in runnable code, separated from the grammar. This could take the form of an interpreter function, or a more object-oriented visitor pattern.

Let us revisit the grammar in \cref{code:gggp_ex} and the corresponding encoding in \cref{code:stgp_ex} to see how these rules can be applied. The non-terminals $Real$ and $Vector$ correspond to abstract classes. The non-terminals $RealLit$ and $d$ are not necessary as they exist only due to the textual nature of BNF, and can be replaced by the native type $float$. Each production results in a concrete class, with a field for each non-terminal included in that production.

\subsection{Meta-Handlers}

Programming languages often have base types (such as integers, floats, strings) that can be complex to implement in BNF syntax. For instance, integers are often represented as \lstinline@<I> ::= <P> | -<P>,  <P> ::= <D><P> | , <D> ::= 0 | 1 | 2 | 3 | 4 | 5 | 6 | 8 | 9@. 

Depending on the language, this approach can lead to int overflowing or lack of floating-point precision. Grammars designers should design strict subsets of the official grammar of the target language, with all the lengthy implementation work that it entails.

Our datatype approach does not require grammar designers to specify the syntactic nature of literal values. Instead, designers define the semantic type of that literal (e.g., \lstinline|int| or \lstinline|float|).

However, it is often useful to further restrict these values. For instance, it might be relevant to restrict integers to be positive, or less than 10. To achieve this, we propose Meta-Handlers, inspired by Refinement Typed Genetic Programming~\cite{DBLP:conf/ppsn/FonsecaSS20}.

Meta-Handlers are type refinements. In the running example, \lstinline|Annotated[int, IntBetween[0,3]]| refines the \lstinline|int| datatype with the Meta-Handler \lstinline|IntBetween[0,3]|. In Refinement Types, this predicate would restrict the int values that belong to this refined type. However, Meta-Handlers serve a dual purpose: they override the tree generation process.

During the initialization phase of GP, a population of individuals is generated from a grammar. In GGGP (like in PonyGE2), non-terminal symbols are iteratively and non-deterministically replaced by valid productions until there is no non-terminal. The final result should be a string in the target language, that will be parsed into an abstract syntax tree to be evaluated, or compiled.

In our proposed encoding, the grammar is used to construct the abstract syntax tree directly (avoiding redundant string operations and parsing). Starting with the non-terminal root type, a concrete subclass is non-deterministically selected, and that object is created. For each field (or constructor argument), a compatible object is constructed by recursively calling this algorithm. The base case is when native types (like \lstinline|int| or \lstinline|float|) are used, and then native random value generators are used.

However, if a type is annotated with a Meta-Handler, instead of calling the recursive algorithm, the tree construction is delegated to that Meta-Handler. For instance, instead of generating a random value in the target language,  \lstinline|Annotated[int, IntBetween[0,3]]| will only generate integers between 0 and 3.
The same could be done for other data types, including polymorphic ones like lists, and even user-defined types if grammar designers also define the corresponding Meta-Handlers.

These overrides can be used for changing the non-deterministic nature of value generation. For instance, instead of generating integers following a uniform distribution, a Meta-Handler could replace that method with taking numbers following a given normal distribution. CFGs, even including probabilistic CFGs~\cite{10.1007/978-3-319-13356-0_1}, could not express this, as the designer would need to define an infinite number of productions. This example proves that Meta-Handlers increase the expressive power of the grammar in GGGP approaches and the types in non-dependent STGP approaches.

Another example where Meta-Handlers contribute to expressive power is through dependent types~\cite{DBLP:conf/ppsn/FonsecaSS20}. Consider the example of generating a pair of integers $(i, j)$ where $j > i$. CFGs do not support this, but a simple Meta-handler could generate any value of $i$ and restrict the minimum value of $j$ to be $i$.

PonyGE2, that uses BNF, supports a custom fixed macro for selecting a value from a list (\texttt{GE\_RANGE:li}), but is limited to that example and, mainly, does not support user-defined macros. It is also important to note that textual macros are not as expressive as type-guided hygienic macros, which would be equivalent to Meta-Handlers.

Attribute Grammars~\cite{DBLP:conf/iwinac/CruzPA05}, where semantic actions are associated with grammar productions are also more powerful than CFG, and equivalent to our encoding with Meta-Handlers. However, Attribute Grammars also suffer from the same issue of interleaving two languages, and work by aborting the tree generation process~\cite{DBLP:conf/iwinac/CruzPA05}, instead of guiding it, like with Meta-Handlers.

\section{Evaluation}
\label{sec:evaluation}

We evaluate our approach against BNF and similar interfaces for grammars in three aspects: feasibility, expressive power, performance, and ergonomics.

\subsection{Feasibility}

To evaluate the feasibility of our design, we implemented this approach in a new Python GP framework, named \GeneticEngine{} that features a data-type frontend, but uses grammars internally. This implementation can be considered a hybrid of GGGP and STGP, and it is open-source and available at \url{https://github.com/alcides/GeneticEngine} and can be installed using \lstinline|pip install geneticengine|. 

To implement the proposed approach, we resorted to type reflection to list abstract classes (to use as non-terminal symbols), and to identify the parameter types of constructors (to instantiate productions). Furthermore, we used the method resolution order (via the \lstinline|mro()| method) to detect which classes were subclasses of other classes. This approach could be implemented in other dynamic languages that support these reflection features, or in a more statically typed language through the use of compile-time macros.

Furthermore, we implemented Meta-Handlers within this framework, to increase the expressive power of our solution.  In particular, \cref{code:metahandlers_examples} presents examples of a few available Meta-Handlers, implemented in pure Python.

\begin{lstfloat}
\begin{lstlisting}[language=Python,caption=Examples of Meta-handlers,label=code:metahandlers_examples]
@dataclass
class Container(Parent):
	age : Annotated[int, IntRange[0, 200]]
	height : Annotated[float, FloatRange[0.10, 2.5]]
	children : Annotated[List[Container], ListSizeBetween[1,30]]
	var_name : Annotated[str, VarRange[["x", "y", "z"]]]
\end{lstlisting}
\end{lstfloat}

We implemented several working examples of GGGP on top of this framework, showing the feasibility of our proposed approach.

\subsection{Expressive Power}

Our grammar encoding method is capable of expressing any CFG, shown by the direct encoding of CFGs according to the rules introduced in \cref{subsec:types_as_grammars}. With Meta-Handlers, we support the semantic actions of Attribute Grammars as long as the programming language is Turing-complete, as are most of the target languages of GGGP. In particular, because tree nodes are built bottom up, our approach is equivalent to S-attributed grammars, in which the evaluation order is well-defined.

While not relevant for the theoretical expressive power, it is important to compare the practical expressive power, in terms of equivalent abstraction levels. BNF grammars can be directly encoded in the type hierarchy, requiring to be as verbose as the target language requires. For instance, if targeting Haskell or Scala, the lines of code required will be equivalent to the BNF syntax.

EBNF~\cite{jtc1996information} is a more convenient syntax than BNF, as it provides some useful syntax sugar features. Optional symbols (\texttt{[a]}) can be achieved in our approach using the \lstinline|Optional| or \lstinline|Maybe| datatype. Repeated symbols (\texttt{a\{1-3\})} can be achieved using  the \lstinline{List} datatype. Finally, static special sequences (\texttt{?? a ??}), which are often not implemented, can be achieved using Meta-Handlers. However, Meta-Handlers are more expressive than static macros in the EBNF syntax.

Furthermore, if considering practical expressive power, our Meta-Handler approach can factor out probabilistic patterns in the language, that a Probabilistic CFG cannot. As a realistic example, consider a grammar that wants to generate programs with a list of DNA motifs. Motifs have an alphabet of four letters (A, T, C, and G) and have a probability for each letter at each position. For instance, the very simple (and artificial) example $[.5, .5, 0, 0], [1, 0, 0, 0]$ would have length 2, and it would have the same probability of generating AA and AT. Each position is represented as the probability of the corresponding letter being in that given position. Probabilistic CFGs can express this pattern (\lstinline|M ::= L1 A, L1 ::= (.5) A, L1 ::= (.5) T |), but require each letter with alternatives to be expressed. Our motif Meta-Handler requires only the motif-defining matrix to be passed, and it handles the generation of trees according to that motif probability. This is very useful when a series of motifs has a higher probability of appearing in the solution than random words.

A common solution to implement probabilistic grammars in GGGP is to use Probabilistic CFG syntax~\cite{10.1007/978-3-319-13356-0_1}, that extends CFG with a probability, or weight, for each production. Our implementation supports the same feature, by annotating each concrete type with a probability (\lstinline|@weight(0.1) class A(S): pass|). 

\subsection{Performance}

We evaluate the performance implications of using a GGGP-STGP hybrid approach (through our \GeneticEngine{} implementation) against a textual BNF approach (through the PonyGE2 implementation). We implemented five examples (Classification task, Regression Task, VectorialGP, String Match, and Game of Life) in both frameworks. We evaluate the performance of both using the same algorithm conditions (initialization method, population size, operator probabilities). The code for replicating these experiments is available at \url{https://github.com/pcanelas/GeneticEngineEvaluation}.


We executed the latest version of both frameworks, using Python 3.9 running on Ubuntu 20.04.1 on an Intel Xeon X5670 24-Core Processor with 24GB of RAM. 

\botafigwide[.8]{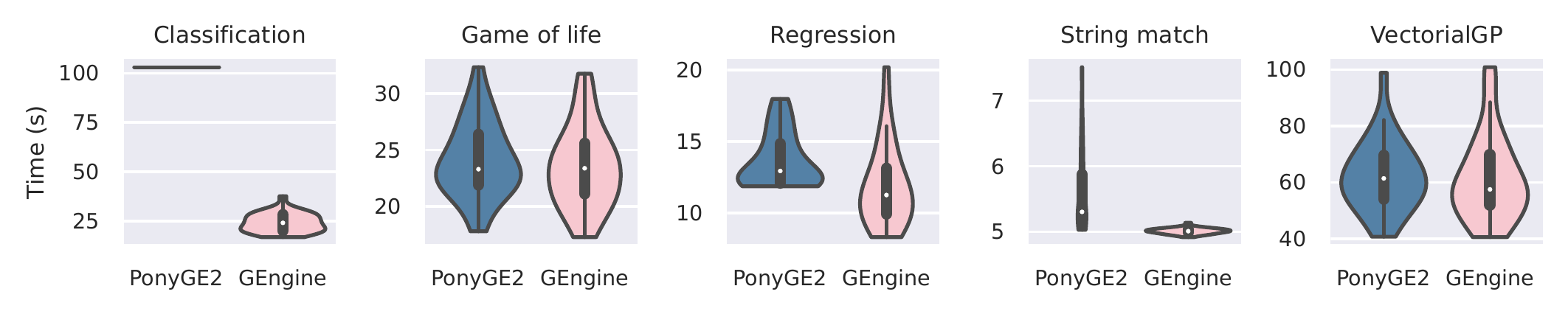}{Execution time, in seconds, of executing PonyGE2 and \GeneticEngine{} on the same benchmarks, under the same conditions (including the number of generations), modulo the differences in the approach. A lower time is better.}

\Cref{fig:merged_plots_time_faceted} shows the distribution of executing each of the five benchmarks with 30 different seeds. Both platforms had the same exact configurations for each benchmark, including a fixed number of generations. The details of each configuration are available in the replication package. The time distribution shows that \GeneticEngine{} is not overall slower than PonyGE2. In fact, in the Classification benchmark, it takes less than half of the time. These results evidence that the ergonomic advantages of \GeneticEngine{} do not come at the cost of performance.

\botafigwide[.8]{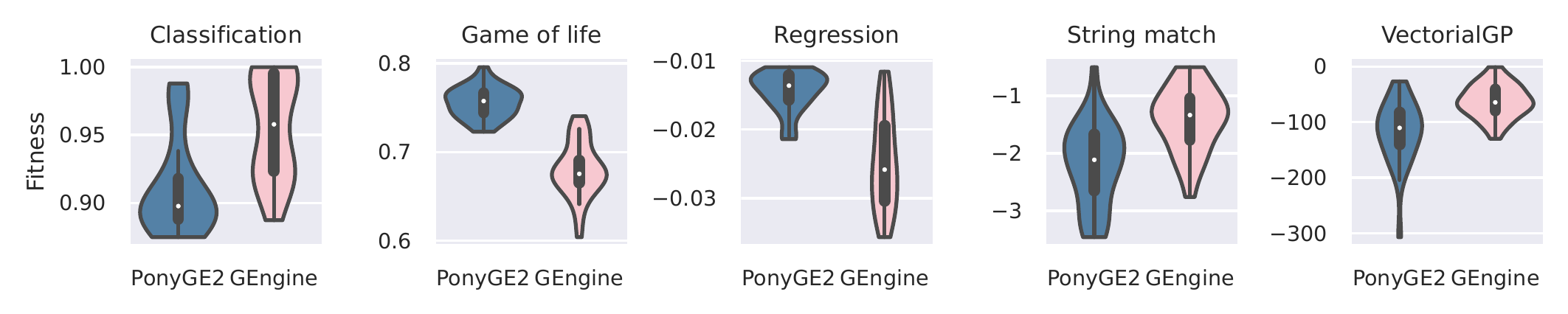}{Fitness comparison of PonyGE2 and \GeneticEngine{} on the same benchmarks, under the same conditions, with a budget of 1 minute. Higher is better.}

\Cref{fig:merged_plots_fitness_faceted} shows the distribution of the fitness of the five benchmarks, run with 30 different seeds. These runs had the same configuration as before, but the stopping criterion was the time budget of 1 minute. Again, it is possible to see that fitness is generally on par with PonyGE2. 

From this evaluation, we can conclude that our encoding does not introduce significant overhead in the process. In fact, avoiding textual representation of programs and subsequent parsing allows our approach to be faster than PonyGE2 in some cases.

\subsection{Ergonomics}

According to the Rust language ergonomics initiative~\cite{turon_2017}, Ergonomics is a measure of the friction you experience when trying to get things done with a tool.

We compare the ergonomics of our approach against BNF and its derivatives using the framework of \citet{DBLP:journals/jsi/Kiper94}, also used to evaluate software development tools. We compare both approaches with regards to Granularity, Coherence and Harmony.

Granularity refers to what is the smallest unit which can have feedback, and it should be as fine as possible. The granularity of our approach is higher than the granularity of existing BNF syntaxes (including those describing Attribute Grammars). A user can only validate their BNF syntax at the file level, by using the grammar in a GGGP evolution run. On the other hand, our approach has a finer granularity, at the token level, because parsers and typecheckers provide immediate feedback on the syntactic and semantic correctness (such as Pylance for Python, or Roslyn for C\#). In this spectrum, our approach is more ergonomic due to the finer granularity level.

Coherence refers to how the units are related and integrated with each other. BNF syntaxes require the user to edit streams of characters, with no notion of structure other than defining productions. As an example, the textual representation of the target language has no meaning within a BNF syntax. Our approach has a higher degree of coherence, as programs can be organized in methods, classes and modules, as in most mainstream programming languages. Furthermore, code editors have semantic visualizations of these structural components and allow semantic actions over them (such as renaming a class across the whole project). This is not possible with textual BNF representations. As such, our approach is also better according to the Coherence criterion.

Harmony refers to agreement between the elements and the corresponding unit, and should be high (as opposed to dissonant). BNF languages are dissonant regarding the two languages in which programmers define their grammars. The granularity of BNF is not harmonious with the granularity of the target language. On the other hand, the granularity of our approach is exactly the same of the target language, showing the advantage of encoding grammars in the target language. Furthermore, error messages are more harmonious in our language, because they are located in the relevant grammar component (grammar or abstract class), while in BNF approaches it is dissonant, as the error will occur in the target language, and not at the grammar level. As such, we can conclude that our approach is more harmonic than BNF-approaches.

We can conclude that our approach is more ergonomic than BNF-based textual approaches, using the framework of \citet{DBLP:journals/jsi/Kiper94}. Furthermore, we identify many concrete advantages practitioners have by using our approach:

\begin{itemize}
	\item \textbf{Type-checked Grammars:} Being part of valid code, grammars are type-checked. Errors in the design of grammar can be immediately detected using a typechecker.
	\item \textbf{Localized Error Messages:} Syntactic and semantic errors (both static and dynamic) include a precise localization of the error, while in textual representations users have to backtrack the location to the BNF location.
	\item \textbf{Autocompletion:} In larger codebases, developers rely on autocompletion tools to avoid unnecessarily navigating files and browsing for the wanted method or variable names. Textual BNF representation prevents these tools from working, but our encoding supports them.
	\item \textbf{Refactoring Tools:} Developers rely on many refactoring actions in their editors to evolve software codebases. Having code inside the textual BNF syntax prevents this tooling support from working. By encoding grammars in the target language, all refactors can be applied to both the grammar and the semantic actions.
	\item \textbf{Testing Infrastructure:} Encoding grammars in the language gives support for designing unit and integration tests that validate the grammar design. This is useful to assure the quality of grammars in real-world scenarios.
	\item \textbf{Auto Formatters:} When working on larger projects, it is important to minimize version control conflicts caused by different formatting choices. By encoding the language in the target language, users can reuse linters for the target language to avoid unnecessary conflicts.
\end{itemize}

\section{Conclusion}
\label{sec:conclusions}

We proposed an approach that replaces BNF grammars as the interface for GGGP (both Grammatical Evolution and Tree-based GP), based on an Object-Oriented type system. As such, it could be considered to provide the "Best of Both Worlds"~\cite{whigham2015examining}. This GGGP-STGP hybrid embeds the grammar within the host language, providing a series of advantages over traditional GGGP: It supports type-checking to validate the consistency and integration of the grammar, autocomplete, auto-imports, can be integrated into existing projects, can be tested and auto-formatted. These advantages translate to better ergonomics, evaluated using the framework of \citet{DBLP:journals/jsi/Kiper94}. These advantages come at no significant performance overhead or lack of expressive power. In fact, the proposal of Meta-Handlers as user-customizable tree-generation overrides provides more expressive power, both from a theoretical perspective (dependent types) and a practical perspective (to refactor grammar patterns).

\begin{acks}
  This work was supported by \textit{Fundação para a Ciência e Tecnologia} (FCT) in the LASIGE Research Unit under the ref. UIDB/00408/2020 and UIDP/00408/2020 and through 
the PhD fellowships under the refs. SFRH/BD/137062/2018 
and UI/BD/151179/2021, 
the CMU--Portugal Dual Degree PhD Program (SFRH/BD/151469/2021), 
by the CMU--Portugal project CAMELOT (POCI-01-0247-FEDER-045915), 
and the RAP project under the reference (EXPL/CCI-COM/1306/2021).
\end{acks}

\bibliographystyle{ACM-Reference-Format}
\bibliography{references}

\end{document}